\documentclass[12pt,letter]{article}

\usepackage{graphicx, epsfig, color}
\def\to{\rightarrow}

\def\bi{\begin{itemize}}
\def\ei{\end{itemize}}

\def\ta{\tilde a}

\def\sps1ap{SPS1a$^\prime$}
\def\c1p{C1$^\prime$}

\def\tG{\tilde G}

\def\tnu{\tilde\nu}

\def\tq{\tilde q}

\def\tz{\widetilde Z}
\def\alt{\stackrel{<}{\sim}}
\def\agt{\stackrel{>}{\sim}}
\def\be{\begin{equation}}  
\def\ee{\end{equation}}  
\def\bea{\begin{eqnarray}}  
\def\eea{\end{eqnarray}}  
\def\beas{\begin{eqnarray*}}  
\def\eeas{\end{eqnarray*}}  
\newcommand\prd[3]{{\it Phys.\ Rev.\ }{\bf D #1} (#2) #3}

\newcommand\prl[3]{{\it Phys.\ Rev.\ Lett.\ }{\bf #1} (#2) #3}
\newcommand\plb[3]{{\it Phys.\ Lett.\ }{\bf B #1} (#2) #3}
\newcommand\jhep[3]{{\it J. High Energy Phys.\ }{\bf #1} (#2) #3}

\newcommand\npb[3]{{\it Nucl.\ Phys.\ }{\bf B #1} (#2) #3}

\newcommand{\hepph}[1]{hep-ph/#1}

\newcommand\ppnp[3]{{\it Prog.\ Part.\ Nucl.\ Phys.}{\bf  #1} (#2) #3}

\begin{document}
\begin{titlepage}
\vspace{0.5cm}
\begin{center}
{\Large\bf Supersymmetry and Dark Matter post LHC8:\\
why we may expect both axion and WIMP detection\footnote{
Talk given at Particle Physics and Cosmology meeting (PPC2013), July 8, 2013, 
Deadwood, South Dakota.}
}\\

\vspace{1.2cm} \renewcommand{\thefootnote}{\fnsymbol{footnote}}
{\large Howard Baer$^1$\footnote[1]{Email: baer@nhn.ou.edu }
}\\ 
\vspace{1.2cm} \renewcommand{\thefootnote}{\arabic{footnote}}
{\it 
$^1$Dept. of Physics and Astronomy,
University of Oklahoma, Norman, OK 73019, USA \\
}

\end{center}

\begin{abstract}
In the post-LHC8 era, it is perceived that what is left of SUSY model parameter space is 
highly finetuned in the EW sector (EWFT). 
We discuss how conventional measures overestimate EWFT in SUSY theory. 
Radiatively-driven natural SUSY (RNS) models maintain the SUSY GUT paradigm with low EWFT at 10\% level, 
but are characterized by light higgsinos $\sim 100-300$ GeV and a thermal underabundance of WIMP dark matter.
Implementing the SUSY DFSZ solution to the strong CP problem explains the small $\mu$ parameter
but indicates dark matter should be comprised mainly of axions with a small admixture of higgsino-like WIMPs. 
While RNS might escape LHC14 searches, we would expect ultimately direct detection of 
both WIMPs and axions. 
An $e^+e^-$ collider with $\sqrt{s}\sim 500-600$ GeV should provide a thorough search
for the predicted light higgsinos.
\end{abstract}
\end{titlepage}


\section{Introduction}

The recent discovery of the Higgs boson $h$ at LHC
is a great triumph and seemingly completes the Standard Model\cite{lhchiggs}.
However, it brings with it a well-known conundrum in that scalar particle masses are
quadratically divergent: $m_h^2|_{phys}=m_h^2|_{tree}+\delta m_h^2$ where 
$\delta m_h^2\sim (c/16\pi^2)\Lambda^2$ and where $\Lambda$ is the high energy cut-off below which
the SM is assumed to be the valid effective field theory. Requiring no large finetuning in $m_h^2$
then implies $\Lambda\alt 1$ TeV. If we wish to extend $\Lambda$ to much higher scales, say those associated
with Grand Unification, then a protective symmetry, supersymmetry or SUSY, is needed.
In SUSY extensions of the SM, then all quadratic divergences cancel, leaving only the more benign log
divergences. SUSY receives indirect support from experiment via 
1. gauge coupling unification, 2. radiative breaking of
EW symmetry via the large top quark mass and 3. the fact that $m_h=125.5\pm 0.5$ GeV falls within the 
narrow window predicted by SUSY theories such as the MSSM\cite{mhiggs}. 

SUSY extensions of the SM also allow for several  viable dark matter candidates while none are 
contained within the SM. An essential requirement for SUSY DM is the existence of $R$-parity
conservation (RPC). RPC is motivated by the need for proton stability: if one allows unfettered 
$R$-parity violation, then the proton would decay in a flash. On the theoretical side, 
RPC is a consequence of $SO(10)$ gauge symmetry, which only allows $matter-matter-Higgs$ couplings
whilst RPV requires $matter-matter-matter$ or $matter-Higgs$ couplings. Of course, $SO(10)$ must
ultimately be broken, but many breaking schemes allow for $R$-parity conservation to survive 
the GUT symmetry breaking.

Most popular amongst the SUSY DM candidates is the lightest neutralino $\tz_1$, a WIMP particle.
Other possibilities include the gravitino $\tG$, the axino $\ta$ in SUSY theories which include 
the PQ solution to the strong $CP$ problem, and the right-hand sneutrino $\tnu_R$.
A left-sneutrino LSP was long ago ruled out by direct WIMP search experiments

\section{Problems with SUSY dark matter}

While the several SUSY DM candidates may seem like an embarrassment of riches, 
each of these is not without problems. In the attractive SUSY see-saw mechanism for generating
neutrino masses, the right sneutrino is expected to be up around the GUT scale; 
one must abandon this approach to accommodate $\tnu_R$ as the LSP. 

In the case of axino, the calculation of axino mass
in gravity mediation finds typically $m_{\ta}\sim m_{3/2}$ where $m_{3/2}$ is the gravitino mass
and which sets the scale for SUSY breaking. It is expected then that the axino should be among the
heavier particles, especially in cases where $m_{3/2}\agt 5$ TeV, which provides a solution to the
cosmological gravitino problem (lighter gravitinos can be produced at large rates in the early
universe and if they are too light, but not the LSP, then their late decays can disrupt the successful
picture of Big BangNucleosynthesis (BBN)\cite{bbn}). Axinos also can be produced thermally in the early universe, 
and typically would produce an overabundance of DM\cite{axthprod}. 
Axino masses $\sim keV-MeV$ size seem  required for the axino to be a viable DM candidate. 
The axino as a DM candidate might be more at home in GMSB
scenarios where the gravitino can also be quite light. 

The gravitino could also be the DM. In this case, SUSY particles produced in the early universe 
would suffer late decays to $\tG$, thus again facing formidable BBN constraints\cite{bbn}.
If the $\tG$ is very light, as in GMSB models, then enhanced decays via its Goldstino component
can accelerate decay rates thus avoiding BBN problems. But the simplest GMSB models seem
highly stressed if not ruled out by a combination of Higgs mass and naturalness constraints\cite{bbm2}.

\begin{figure}
\includegraphics[angle=-90,width=12cm]{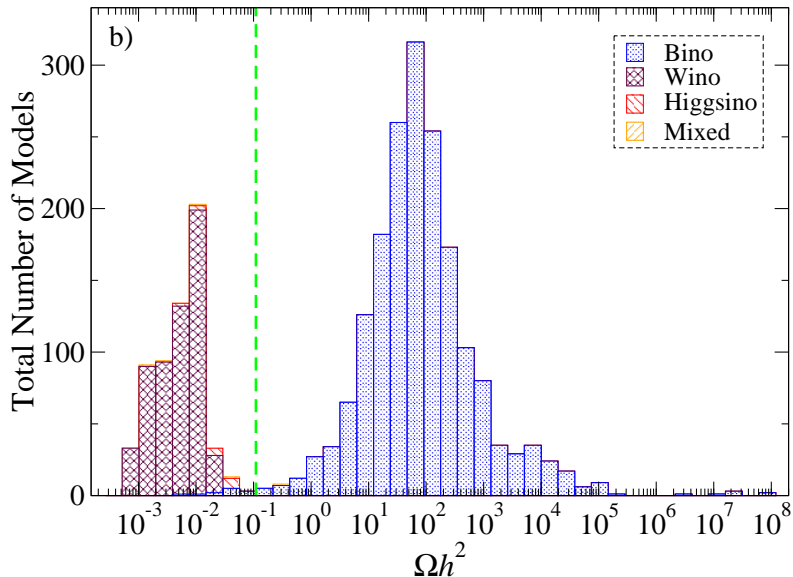}
\caption{Predicted neutralino relic abundance from a scan 
over the 19-parameter SUGRA model, 
while requiring $m_{\tz_1}<500$ GeV (from Ref. \cite{ax19}).
\label{fig:bbs}}
\end{figure}

The lightest neutralino $\tz_1$, as remarked above, is perhaps the most popular SUSY DM candidate.
However, it also is problematic. First, in spite of much hype about a WIMP miracle, 
calculations of $\Omega_{\tz_1}h^2$ in the simple thermal WIMP production scenario imply
typically a large overproduction of DM (typically by several orders of magnitude) 
in the case where $\tz_1$ is bino-like, and underproduction (typically by 1-2 orders of magnitude)
in the case of a higgsino- or wino-like LSP: see Fig. \ref{fig:bbs}. 
To match the measured abundance of CDM, either a 
fine adjustment of relative higgsino-bino-wino components is required\cite{wtn}, or else special
annihilation mechanisms-- via resonance annihilation or co-annihilation-- are needed.
Thus, rather contrived scenarios are required to bring the predicted abundance of 
neutralino dark matter into accord with measurements.

A second problem-- often ignored by SUSY DM practitioners-- is the strong $CP$ problem. 
The PQWW axion solution still seems the most compelling way to address this\cite{pqww}. 
But then the axion is also a viable DM candidate. If the $a$ exists, then the dark matter calculus
undergoes a radical change.

A final challenge to SUSY DM is that recent negative LHC sparticle search results, along with the
rather large value of $m_h$, has led many to speculate that the simple MSSM effective theory is finetuned
in the EW sector: then it may not be the whole picture and perhaps radical new SUSY model 
building ideas are needed\cite{craig}.
This latter viewpoint is mainly fueled by the perception that large logarithms of the form\cite{kn}
\begin{equation}
\delta m_h^2\sim -(3f_t^2/8\pi^2) (m_{Q_3}^2+m_{U_3}^2+A_t^2)\ln (\Lambda^2/m_{SUSY}^2)
\label{eq:log}
\end{equation}
lead to large finetuning in $m_h^2$, in a similar vein as do the quadratic divergences for
the SM Higgs squared mass.

\section{An improved picture of SUSY dark matter: 
mixture of axions and higgsinos}

\subsection{Radiatively-driven natural SUSY}

The log term in Eq. \ref{eq:log} taken at face value seems to imply that-- 
for better than 10\% EWFT and $\Lambda$ as high as $m_{GUT}$--  
SUSY naturalness requires top squarks lighter than 200 GeV and consequently gluinos less 
than about 600 GeV, almost certainly in violation of LHC search constraints. 
However, as emphasized in Ref.~\cite{comp}, the term in Eq. \ref{eq:log} is only one piece of
various {\it non-independent} terms contributing to $m_h^2$. For instance, while
$m_h^2|_{tree}$ and $\delta m_h^2$ are independent in the SM, in the MSSM the
$m_{H_u}^2(\Lambda )$ and $\delta m_{H_u}^2$ are {\it dependent}: the larger one makes
$m_{H_u}^2(\Lambda )$, the larger becomes $\delta m_{H_u}^2$. This suggests that one 
ought to collect $m_{H_u}^2(\Lambda )+\delta m_{H_u}^2$ into a single term, as is done with the
Barbieri-Giudice\cite{bg} measure $\Delta_{BG}$ and the electroweak measure $\Delta_{EW}$\cite{rns,perel}.

While the measure $\Delta_{BG}$ avoids the pitfall of Eq. \ref{eq:log} by re-writing the {\it combination} 
$(m_{H_u}^2(\Lambda )+\delta m_{H_u}^2)$ in terms of fundamental input 
parameters, it is itself highly model-dependent since by definition it measures
fractional change in $m_Z^2$ against fractional change in model parameters.
This means $\Delta_{BG}$ changes from model to model, even if each model generates exactly 
the same weak scale spectrum. As an example, in the focus point region of
mSUGRA, where $m_{H_u}^2\equiv m_{H_d}^2\equiv m_{16}^2(3)$, then large cancellations
due to correlated high scale soft terms yield much lower finetuning than is expected in
more general models. 
In an ultimate theory (UTH) where parameters $A_0$, $m_{1/2}$ and $m_0$ are also expected to be correlated,
then even greater reductions in $\Delta_{BG}$ can occur. 
The lesson here is that the popular effective theories which we use, 
where the high scale soft term values parametrize our ignorance of SUSY breaking, will often 
yield much more finetuning than in more correlated models with fewer free parameters.
What we are really interested in is whether-or-not {\it nature} is EW-finetuned 
(and by implication the UTH which describes it), 
and {\it not} how finetuned are the more general effective theories which might contain the UTH\cite{comp}. 

A model-independent measure is given by $\Delta_{EW}$\cite{rns}, which measures {\it weak scale} SUSY contributions to $m_Z^2$. 
The $\Delta_{EW}$ parameter has been interpreted as a bound on finetuning, 
and as a necessary-- albeit not sufficient-- condition for low EWFT\cite{wp1}. 
Models with low $\Delta_{EW}$ are characterized by 
\begin{itemize}
\item low $|\mu|\sim 100-300$ GeV,
\item $m_{H_u}^2$ is driven radiatively to just small negative values $\sim -m_Z^2$, and
\item top squarks are still bounded, but now in the few TeV-regime. They are also highly mixed. 
The large mixing reduces the radiative corrections to $m_Z^2$ whilst lifting $m_h$ into the 125 GeV regime.
\end{itemize}
Such low $\Delta_{EW}$ models are referred to as radiatively-driven natural SUSY, or RNS, since
$m_{H_u}^2$ is radiatively driven to just small negative values, thus allowing for
EW naturalness. The popular mSUGRA/CMSSM model admits a minimum value of $\Delta_{EW}\sim 200$ 
so is surely finetuned. However, the NUHM2 model allows for small $\mu$ but also
relatively light TeV-scale stops: in this case, $\Delta_{EW}$ values as low as $\sim 7$ can be found. 

In such low $\Delta_{EW}\alt 10$ models, the lightest SUSY particle is typically
the higgsino-like $\tz_1$, albeit with a not-too-small gaugino component.
The thermal relic density of such higgsinos is $\Omega_{\tz_1}h^2\sim 0.005-0.01$, 
{\it i.e.} a factor 10-15 below measured values. 

\subsection{Strong $CP$ problem}

Another possibility for finetuning occurs in the QCD sector. To implement 't~Hooft's solution
to the $U(1)_A$ problem ({\it i.e.}~why there are three and not four light pions),
the term
\be
\frac{\bar{\theta}}{32\pi^2}F_{A\mu\nu}\tilde{F}_A^{\mu\nu}
\label{eq:FF}
\ee
should occur in the QCD Lagrangian, where $\bar{\theta}=\theta +{\rm arg}({\rm det}{\cal M})$,
${\cal M}$ is the quark mass matrix, $F_{A\mu\nu}$ is the gluon field strength and
$\tilde{F}_A^{\mu\nu}$ is its dual.
Measurements of the neutron electric dipole moment (EDM) require $\bar{\theta}\alt 10^{-10}$
so that one might require an enormous cancellation within $\bar{\theta}$~\cite{axreview}. Alternatively,
the PQWW solution~\cite{pqww} introduces an {\it axion} field $a$; the additional axion contributions
allow for the coefficient in Eq. \ref{eq:FF} to dynamically settle to zero, thus solving
the so-called strong $CP$ problem.

In SUSY theories, the axion enters as but one element of an axion {\it superfield} which
necessarily contains also a spin-0 $R$-parity even saxion $s$ and a spin-$1/2$ $R$-parity-odd
axino $\ta$. Calculations of the saxion and axino masses within the context of supergravity~\cite{maxino}
imply $m_s\sim m_{\ta}\sim m_{3/2}$ where the gravitino mass $m_{3/2}$ is expected to be of
order the TeV scale.
If the lightest neutralino ({\it e.g.} the higgsino $\tz_1$) is the lightest SUSY particle (LSP)
in $R$-parity conserving theories, then one would expect dark matter to be comprised of {\it two particles}:
the axion as well as the higgsino-like SUSY WIMP.
The saxion and axino couplings to matter are suppressed by the PQ breaking scale
$f_a$ which may range from $f_a\sim 10^9-10^{16}$ GeV~\cite{axreview}.
While the saxion and axino are expected
to play little or no role in terrestrial experiments, they can have an enormous impact on
dark matter production in the early universe.

\subsection{SUSY DFSZ model and Kim-Nilles solution to the $\mu$ problem}

The PQ symmetry required to solve the strong $CP$ problem can be implemented in two ways.
In the SUSY KSVZ model~\cite{ksvz,kim}, the axion superfield couples to exotic heavy quark/squark superfields
$Q$ and $\bar{Q}$ which carry PQ charges. 
Alternatively to SUSY KSVZ, in the SUSY DFSZ model~\cite{dfsz,kimnilles,chun,bci} the
PQ superfield couples directly to the Higgs superfields carrying non-trivial PQ charges:
\be \label{WDFSZ}
W_{\rm DFSZ}\ni \lambda \frac{S^2}{M_P} H_u H_d .
\ee
Here, $S$ is a Minimal Supersymmetric Standard Model (MSSM) singlet but carries
a PQ charge and contains the axion field.
An advantage of this approach is that it provides a solution to the $\mu$ problem~\cite{kimnilles}:
since the $\mu$ term is supersymmetric, one expects $\mu\sim M_P$ in contrast to
phenomenology  which requires $\mu\sim m_{\rm weak}$.
In this Kim-Nilles solution, PQ charge assignments forbid the usual superpotential $\mu$ term.
Upon breaking of PQ symmetry, the field $S$ receives a vev $\langle S\rangle\sim f_a$, so that an
effective $\mu$ term is generated with $\mu\sim \lambda f_a^2/M_P\sim \lambda m_{3/2}$.
For small $\lambda$ one may generate $\mu\sim 100-200$ GeV in accord with naturalness whilst
$m_{\tq}\sim m_{3/2}\sim 10$ TeV in accord with LHC constraints and in accord with at least a partial
decoupling solution to the SUSY flavor, $CP$ and gravitino problems~\cite{dine}.

\subsection{Axino and saxion production/decay in SUSY DFSZ}

In the SUSY KSVZ model, the derivative coupling of axions to matter leads to thermal production rates
for axinos and saxions which are proportional to the re-heat temperature $T_R$ after inflation.
In contrast, in SUSY DFSZ the direct coupling of axion superfield to higgs superfields leads to production
rates independent of $T_R$. In addition, as usual, saxions can be produced via coherent saxion field oscillations
which are important for large saxion field strength $s_i$ which is assumed comparable to $f_a$.
Also, axions are produced as usual via coherent oscillations/vacuum mis-alignment 
around the QCD phase transition.

Once produced, then axinos decay (in SUSY DFSZ) into mainly higgsino plus Higgs or higgsino plus 
gauge boson states. The decays are quicker in SUSY DFSZ than in SUSY KSVZ, and for $f_a\alt 10^{12}$ GeV, they occur
before neutralino freeze-out. 
Likewise, saxions decay dominantly to higgsino pairs (leading to additional WIMP production) 
or to axion pairs when the $\xi$ parameter $\sim 1$ (leading to dark radiation)\cite{darkrad}\footnote{Here, $\xi \equiv \sum_i q_i^3v_i^2/f_a^2$. Later,
$\xi =\Omega_{\tz_1}^{std}h^2/0.12$.}.
For $f_a\alt 10^{12}$ GeV, they tend to decay before WIMP freeze-out, so the simple thermal WIMP
production rate should remain valid. The dark radiation from saxion decay is slight for $f_a\alt 10^{14}$ GeV.

\subsection{Axion and higgsino relic density}

\begin{figure}
  \includegraphics[height=.3\textheight]{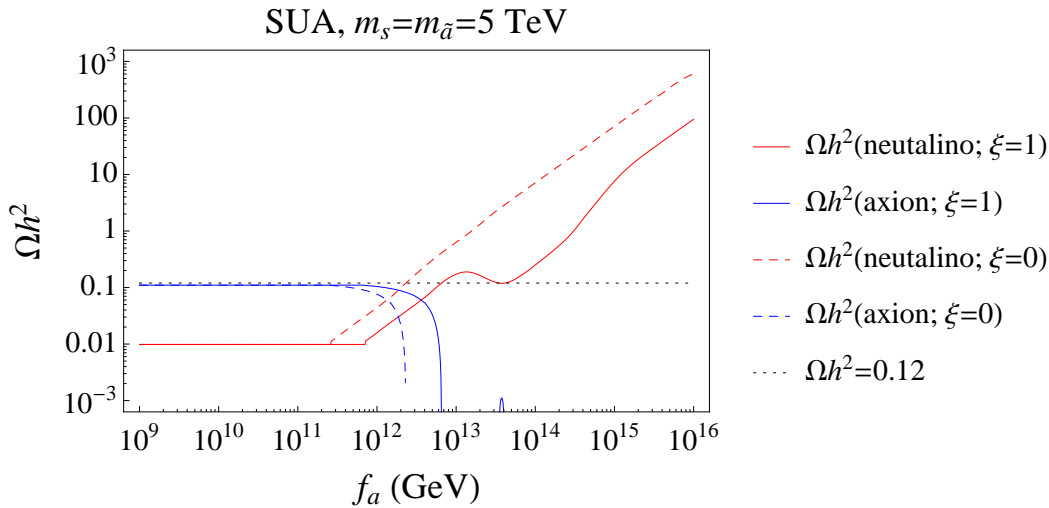}
  \caption{Neutralino and axion relic abundance from the SUSY DFSZ
axion model versus PW scale $f_a$ for the SUA benchmark point.
\label{fig:Oh2}}
\end{figure}

\begin{figure}
  \includegraphics[height=.3\textheight]{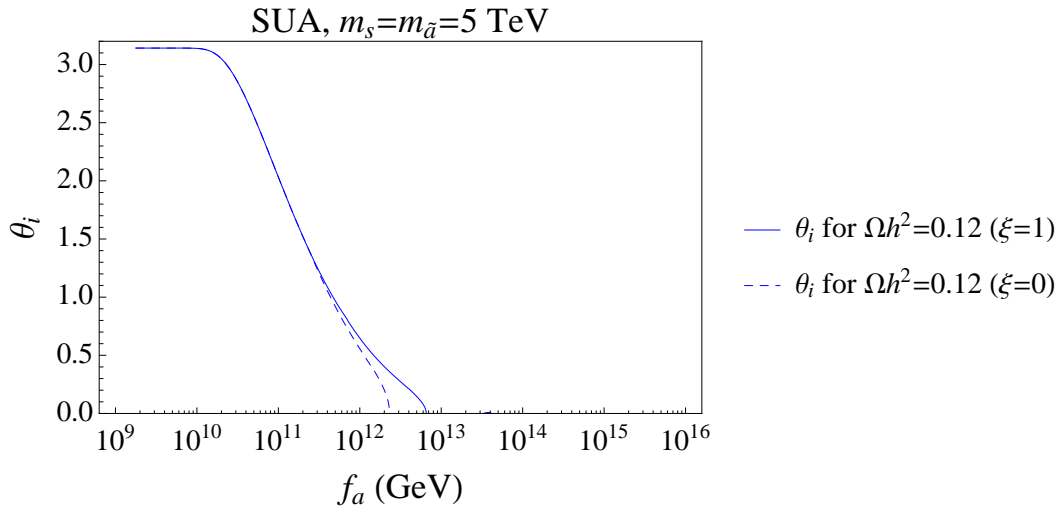}
  \caption{Axion field misalignment angle vs. $f_a$
which is required to saturate mixed axion-neutralino abundance for
$\xi=0$ (dashed) and $\xi=1$  (solid).
\label{fig:theta}}
\end{figure}

Let us now examine the contributions of neutralinos and axions to the observed dark matter density
expected in the SUSY DFSZ model. 
Our result is shown in Fig.~\ref{fig:Oh2} assuming $m_{\ta}=m_s =5$ TeV\cite{axltr}.
Starting at $f_a =10^9$ GeV as required by astrophysical constraints, the neutralino
abundance $\Omega_{\tz_1}h^2\approx 0.01$ is given by the standard thermal freeze-out over a {\it large range}
of $f_a$ extending all the way up to $f_a\sim 10^{12}$ GeV.
In this regime, the axion abundance can always be found by adjusting $\theta_i$
such that the summed abundance meets the measured value: $\Omega_{\tz_1}h^2+\Omega_ah^2 =0.12$.
The required value of $\theta_i$ is shown in Fig.~\ref{fig:theta}. For very low $f_a\sim 10^9$ GeV, a large
value of $\theta_i\sim \pi$ is required, and $\Omega_a h^2$ is dominated by the anharmonicity term.
As $f_a$ increases, the assumed initial axion field value $\theta_i f_a$ increases, so the required
misalignment angle $\theta_i$ decreases. Values of $\theta_i\sim 1$ are found around $f_a\sim 2\times 10^{11}$ GeV
for both $\xi =0$ and 1. In this entire region with $f_a\sim 10^9-10^{12}$ GeV, we expect from natural SUSY
that the relic higgsino abundance lies at the standard freezeout value, comprising about 5-10\% of the total
dark matter density, while axions would comprise 90-95\% of the abundance.
Thus, over the commonly considered range of $f_a$, we expect
{\it mainly axion cold dark matter from natural SUSY},
along with a non-negligible fraction of higgsino-like WIMPs.

\subsection{Detection of axions and higgsinos}

\begin{figure}
  \includegraphics[height=.35\textheight]{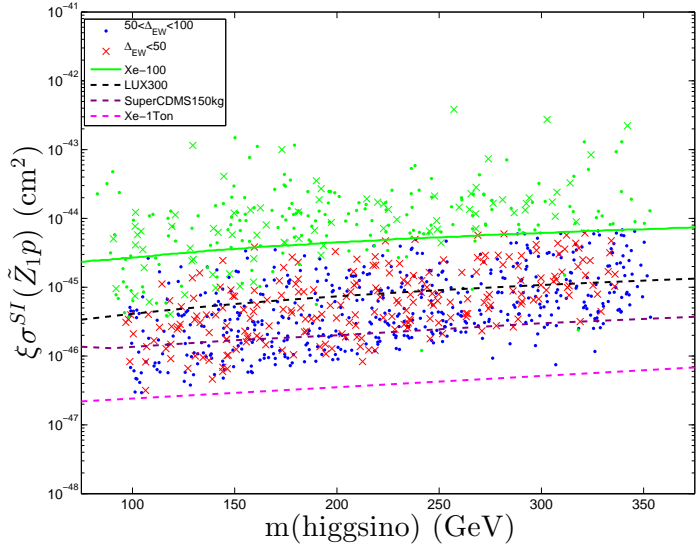}
  \caption{Plot of rescaled higgsino-like WIMP spin-independent 
direct detection rate $\xi \sigma^{SI}(\tz_1 p)$ 
versus $m(higgsino)$ from a scan over NUHM2 parameter space with $\Delta_{EW}<50$ (red crosses)
and $\Delta_{EW}<100$ (blue dots). 
Green points are excluded by current direct/indirect WIMP search experiments.
We also show the current reach from $Xe$-100 experiment, 
and projected reaches of LUX, SuperCDMS 150 kg and $Xe$-1 ton.
\label{fig:SI}}
\end{figure}

In Fig.~\ref{fig:SI}, we show the spin-independent higgsino-proton scattering rate in $pb$
as calculated using IsaReS\cite{IsaReS}. The result is rescaled by a factor $\xi=\Omega_{\tz_1}^{std}h^2/0.11$
to account for the fact that the local relic abundance might be far less than the usually assumed value
$\rho_{local}\simeq 0.3$ GeV/cm$^3$, as suggested long ago by Bottino {\it et al.}\cite{bottino} 
(the remainder would be composed of axions). 
The higgsino-like WIMP in our case scatters from quarks and gluons mainly via $h$ exchange. 
The $\tz_1 -\tz_1 -h$ coupling involves a product of both higgsino and gaugino components. In the case of RNS models, 
the $\tz_1$ is mainly higgsino-like, but since $m_{1/2}$ is bounded from above by naturalness, the $\tz_1$
contains enough gaugino component that the coupling is never small:  in the notation of Ref. \cite{wss}
\be
{\cal L}\ni -X_{11}^h \overline{\tz}_1 \tz_1 h
\ee
where
\be
X_{11}^h =-{1\over 2}\left(v_2^{(1)}\sin\alpha -v_1^{(1)}\cos\alpha \right) 
\left(gv_3^{(1)}-g'v_4^{(1)}\right) ,
\ee
and where $v_1^{(1)}$ and $v_2^{(1)}$ are the higgsino components and $v_3^{(1)}$ and $v_4^{(1)}$ are
the gaugino components of the lightest neutralino, $\alpha$ is the Higgs mixing angle and $g$ and 
$g^\prime$ are $SU(2)_L$ and $U(1)_Y$ gauge couplings.
Thus, for SUSY models with low $\Delta_{EW}\alt 50-100$, the SI direct detection cross section is also 
bounded from below, even including the rescaling factor $\xi$.

From the Figure, we see that the current reach from 225 live-days
of Xe-100 running\cite{xe100} already bites into a significant spread of parameter points.
The excluded points are colored green.
The projected reach of the LUX 300 kg detector\cite{lux} is also shown by the black-dashed contour, which
should explore roughly half the allowed RNS points. 
The reach of SuperCDMS 150 kg detector\cite{cdms} is shown as the purple-dashed contour.
The projected reach of Xe-1-ton, a ton scale
liquid Xenon detector, is also shown\cite{xe1ton}. A major result is this: the projected Xe-1-ton detector--
or other comparable WIMP direct-detectors-- can make a {\it complete}
exploration of the RNS parameter space. Since deployment of the Xe-1-ton detector is imminent,
it seems direct WIMP search experiments may either verify or exclude RNS models in the near future,
thus bringing the story of electroweak naturalness to a conclusion!

\begin{figure}
  \includegraphics[height=.35\textheight]{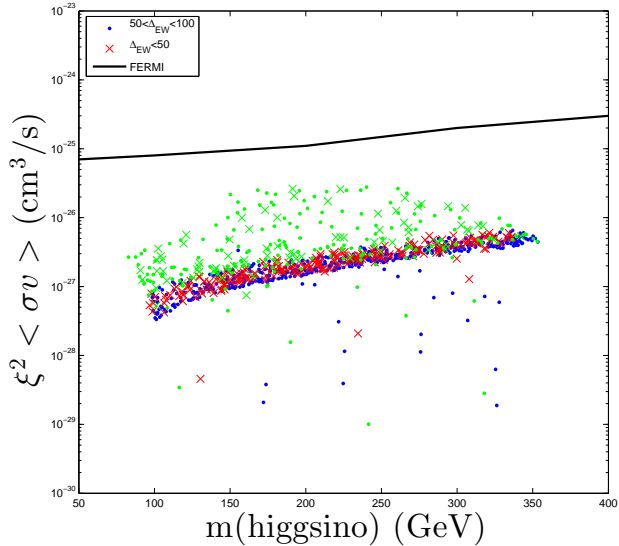}
  \caption{Plot of rescaled $\xi^2 \langle\sigma v\rangle |_{v\to 0}$ 
versus $m(higgsino)$ from a scan over NUHM2 parameter space with $\Delta_{EW}<50$ (red crosses)
and $\Delta_{EW}<100$ (blue dots). 
Green points are excluded by current direct/indirect WIMP search experiments.
We also show current reach from Fermi LAT, 
Ref. \cite{fermi}.
\label{fig:sigv}}
\end{figure}

In Fig.~\ref{fig:sigv}, we show the rescaled thermally-averaged neutralino annihilation
cross section times relative velocity in the limit as $v\to 0$: $\xi^2\langle\sigma v\rangle|_{v\to 0}$.
This quantity enters into the rate expected from WIMP halo annihilations into
$\gamma$, $e^+$, $\bar{p}$ or $\bar{d}$. 
The rescaling appears as $\xi^2$ since limits depend on the square of the local WIMP abundance\cite{bottino_id}.
Anomalies in the positron and $\gamma$ spectra
have been reported, although the former may be attributed to pulsars\cite{pulsars}, 
while the latter 130 GeV gamma line may be instrumental. 
On the plot, we show the limit derived from 
the Fermi LAT gamma ray observatory\cite{fermi} for WIMP annihilations into $WW$. 
These limits have not yet reached the RNS parameter space due in part to the squared rescaling factor.

\section{Conclusions}

 Supersymmetry with not too heavy top squarks, low higgsino mass $\mu\sim 100-200$ GeV
and PQWW solution to the strong CP problem successfully avoids high finetuning in both the EW and QCD sectors of the theory
while evading LHC constraints. The SUSY DFSZ model, wherein Higgs superfields carry PQ charge, also provides a solution to
the SUSY $\mu$ problem. In such models, over a large range of PQ breaking scale $f_a\sim 10^9-10^{12}$ GeV,
saxions and axinos typically decay before neutralino freeze-out so that the higgsino portion of dark matter is
expected to lie in the 5-10\% range while axions would comprise the remainder: 90-95\%. The relic higgsinos
ought to be detectable at ton scale noble liquid detectors, even with a depleted local abundance, while
indirect detection should be more limited since expected rates go as the depleted abundance squared~\cite{bbm}.
Prospects are bright for microwave cavity detection of axions since the range of $f_a$ where mainly axion
dark matter is expected should be accessible to experimental searches~\cite{axsearch}.
While corroborative searches for natural SUSY with light higgsinos is limited at the LHC~\cite{wp},
a definitive higgsino search should be possible at
$e^+e^-$ colliders with $\sqrt{s}$ up to $500-600$ GeV.

\section*{Acknowledgments}

I thank the organizers of PPC 2013 for providing me with the opportunity to give this talk.
I also thank my collaborators K. J. Bae, E. J. Chun, A. Lessa, V. Barger, P. Huang, D. Mickelson, A. Mustafayev,
W. Sreethawong and X. Tata for their excellent work on these issues.

\end{document}